\documentclass[journal]{IEEEtran}
\usepackage{setspace}
\usepackage{graphicx}
\usepackage{booktabs}
\usepackage[final]{changes} %clear traces of changes for submission
\ifCLASSINFOpdf
\else
\fi
\usepackage{amsmath}
\usepackage{amssymb}

\begin{document}
%
% paper title
% Titles are generally capitalized except for words such as a, an, and, as,
% at, but, by, for, in, nor, of, on, or, the, to and up, which are usually
% not capitalized unless they are the first or last word of the title.
% Linebreaks \\ can be used within to get better formatting as desired.
% Do not put math or special symbols in the title.
\title{Microfabricated Neuroaccelerometer: Integrating Sensing and Reservoir Computing in MEMS}
%
%
% author names and IEEE memberships
% note positions of commas and nonbreaking spaces ( ~ ) LaTeX will not break
% a structure at a ~ so this keeps an author's name from being broken across
% two lines.
% use \thanks{} to gain access to the first footnote area
% a separate \thanks must be used for each paragraph as LaTeX2e's \thanks
% was not built to handle multiple paragraphs
%

\author{Bruno~Barazani, %~\IEEEmembership{Member,~IEEE,}
        Guillaume~Dion, %~\IEEEmembership{Fellow,~OSA,}
        Jean-François~Morissette, %~\IEEEmembership{Fellow,~OSA,}
        Louis~Beaudoin, %~\IEEEmembership{Fellow,~OSA,}
        and~Julien~Sylvestre%~\IEEEmembership{Life~Fellow,~IEEE}% <-this % stops a space
\thanks{B. Barazani, G. Dion, J.-F. Morissette, L. Beaudoin, and J. Sylvestre are with Interdisciplinary Institute for Technological Innovation - 3IT at University of Sherbrooke, 3000 boul. de l'Université J1K 0A5 Sherbrooke}% <-this % stops a space
\thanks{© 2020 IEEE.  Personal use of this material is permitted.  Permission from IEEE must be obtained for all other uses, in any current or future media, including reprinting/republishing this material for advertising or promotional purposes, creating new collective works, for resale or redistribution to servers or lists, or reuse of any copyrighted component of this work in other works.}
%\thanks{J. Doe and J. Doe are with Anonymous University.}% <-this % stops a space
%\thanks{Manuscript received April 19, 2005; revised August 26, 2015.}
}

% note the % following the last \IEEEmembership and also \thanks - 
% these prevent an unwanted space from occurring between the last author name
% and the end of the author line. i.e., if you had this:
% 
% \author{....lastname \thanks{...} \thanks{...} }
%                     ^------------^------------^----Do not want these spaces!
%
% a space would be appended to the last name and could cause every name on that
% line to be shifted left slightly. This is one of those "LaTeX things". For
% instance, "\textbf{A} \textbf{B}" will typeset as "A B" not "AB". To get
% "AB" then you have to do: "\textbf{A}\textbf{B}"
% \thanks is no different in this regard, so shield the last } of each \thanks
% that ends a line with a % and do not let a space in before the next \thanks.
% Spaces after \IEEEmembership other than the last one are OK (and needed) as
% you are supposed to have spaces between the names. For what it is worth,
% this is a minor point as most people would not even notice if the said evil
% space somehow managed to creep in.

% The paper headers
\markboth{Accepted by the Journal of Microelectromechanical Systems}%
{Shell \MakeLowercase{\textit{Barazani et al.}}: Microfabricated Neuroaccelerometer}
% The only time the second header will appear is for the odd numbered pages
% after the title page when using the twoside option.
% 
% *** Note that you probably will NOT want to include the author's ***
% *** name in the headers of peer review papers.                   ***
% You can use \ifCLASSOPTIONpeerreview for conditional compilation here if
% you desire.

% If you want to put a publisher's ID mark on the page you can do it like
% this:
%\IEEEpubid{0000--0000/00\$00.00~\copyright~2015 IEEE}
% Remember, if you use this you must call \IEEEpubidadjcol in the second
% column for its text to clear the IEEEpubid mark.

% use for special paper notices
%\IEEEspecialpapernotice{(Invited Paper)}

% make the title area
\maketitle

% As a general rule, do not put math, special symbols or citations
% in the abstract or keywords.
\begin{abstract}
This study presents the design, fabrication, and test of a micro accelerometer with intrinsic processing capabilities, that integrates the functions of sensing and computing in the same MEMS.
The device consists of an inertial mass electrostatically coupled to an oscillating beam through a gap of 8 $\mu$m.
The motion of the inertial mass modulates an AC electrostatic field that drives the beam in its non-linear regime.
This non-linearity is used to implement machine learning in the mechanical domain, using reservoir computing with delayed feedback to process the acceleration information provided by the inertial mass.
The device is microfabricated on a silicon-on-insulator substrate using conventional MEMS processes.
Dynamic characterization showed good accelerometer functionalities, with an inertial mass sensitivity on the order of 100 mV/g from 250 to 1300 Hz and a natural frequency of 1.7 kHz.
In order to test the device computing capabilities, two different machine learning benchmarks were implemented, with the inputs fed to the device as accelerations.
The neuromorphic MEMS accelerometer was able to accurately emulate non-linear autoregressive moving average models and compute the parity of random bit streams.
These results were obtained in a test system with a non-trivial transfer function, showing a robustness that is well-suited to anticipated applications.
\end{abstract}

% Note that keywords are not normally used for peerreview papers.

% For peer review papers, you can put extra information on the cover
% page as needed:
% \ifCLASSOPTIONpeerreview
% \begin{center} \bfseries EDICS Category: 3-BBND \end{center}
% \fi
%
% For peerreview papers, this IEEEtran command inserts a page break and
% creates the second title. It will be ignored for other modes.
\IEEEpeerreviewmaketitle

\section{Introduction}
% The very first letter is a 2 line initial drop letter followed
% by the rest of the first word in caps.
% 
% form to use if the first word consists of a single letter:
% \IEEEPARstart{A}{demo} file is ....
% 
% form to use if you need the single drop letter followed by
% normal text (unknown if ever used by the IEEE):
% \IEEEPARstart{A}{}demo file is ....
% 
% Some journals put the first two words in caps:
% \IEEEPARstart{T}{his demo} file is ....
% 
% Here we have the typical use of a "T" for an initial drop letter
% and "HIS" in caps to complete the first word.
%\IEEEPARstart{T}{his} demo file is intended to serve as a ``starter file''
%for IEEE journal papers produced under \LaTeX\ using
%IEEEtran.cls version 1.8b and later.
% You must have at least 2 lines in the paragraph with the drop letter
% (should never be an issue)
%I wish you the best of success.

Control systems are generally built from three classes of devices: sensors convert stimuli into electrical signals, which are processed by an electronic controller, which itself generates control signals sent to actuators.
MEMS technologies are especially popular to implement sensors for different types of stimuli, such as acceleration, pressure (including sound), spatial orientation and temperature.
This popularity stems from low manufacturing costs, as well as from the sensitivity and energetic efficiency of the MEMS sensors.
These benefits are all related to the small physical dimensions of the mechanical components of the MEMS, and the resulting fast dynamics and low mechanical losses.
MEMS sensors thus produce signals that are rich in information about the state of a system; these signals must be properly processed by a controller device, which often must implement complex control strategies.
Familiar examples include underactuated robotic systems (such as quadrotor drones), with controllers generating control signals that are complex functions of the system dynamics and of sensor data (provided by accelerometers or gyroscopes, for instance), as well as anomaly detectors in preventive maintenance systems, where specific vibration patterns must be identified in sensor data (provided by microphones or accelerometers, for instance), often in the presence of colored or non-stationary noise.
This complexity results in controllers which are often much larger and less energy-efficient than the sensor devices in the control system.

In an attempt to develop more efficient control systems through integration, we propose a new class of MEMS devices, where both the sensory {\em and} the computing functions are implemented in the mechanical response of the same device.
While sensory functions are implemented using fairly conventional MEMS designs, computing functions exploit the non-linear dynamics of a mechanical resonator in the MEMS, to implement a form of machine learning known as {\em reservoir computing} (RC) \cite{jaeger_harnessing_2004}.
The implementation of RC in new substrates has been the target of several recent studies, which were able to emulate RCs in different hardware platforms such as memristors  arrays \cite{du_reservoir_2017}, optical systems \cite{paquot_optoelectronic_2012,larger_photonic_2012,vandoorne_experimental_2014}, mechanical devices \cite{degrave_developing_2015}, and spintronic devices \cite{torrejon_neuromorphic_2017}.
This is because new unconventional computing architectures are expected to exceed the density and the \added{energy} efficiency of \deleted{the} current technology.
The approach used in this study enables complex computing to be implemented in the MEMS in a {\em trainable} manner, by the repetitive presentation to the device of appropriate responses to randomly selected sensory inputs.
As the computing process is similar to data processing by a neural network, we call this new class of devices {\em neuromorphic MEMS}.
We here demonstrate experimentally a neuromorphic MEMS accelerometer, or neuroaccelerometer for brevity, by training it to perform two different machine learning benchmark tasks on signals that are applied as physical accelerations on the device.
To the best of our knowledge, this constitutes the first demonstration of a single physical device that is both a sensor and a (neuromorphic) computer.

The design of the neuroaccelerometer is based on a conventional suspended proof mass, that is coupled electrostatically to a beam clamped at both ends (section \ref{sec:design}).
The motion of the proof mass, induced by accelerations applied on the device, modulates the amplitude of a pump signal driving the beam near resonance, thus establishing the coupling between the sensing and the computing portion of the device.
The non-linear dynamics of the beam are exploited to implement a reservoir computer, using a scheme similar to reference \cite{dion_reservoir_2018}, where inputs were applied directly as electrical signals on a similar beam, to demonstrate experimentally that MEMS resonators could be used to perform complex computations, including the classification of spoken words.
While, theoretically, networks of mechanically coupled MEMS resonators could address much more complicated computing tasks \cite{coulombe_computing_2017}, we initially focus on a single resonator for this first demonstration of a neuromorphic MEMS sensor.
As described in section \ref{sec:fabrication}, we have used standard microfabrication techniques to build a neuroaccelerometer.
The mechanical characterization of the neuroaccelerometer is presented in section \ref{sec:charac}, where fairly conventional sensing capabilities are demonstrated.
In section \ref{sec:processing}, the neuromorphic computing capabilities of the device are demonstrated, with machine learning benchmarks realized on input data provided as accelerations acting on the device.
The benchmarks include a first task (NARMA) with requirements that are similar to those of a robotic controller, as well as a second task (Parity) with requirements resembling those of a signal classification controller.
Both benchmarks demand significant non-linear processing and the ability to memorize data for some period of time.
They are also realized on actual hardware, including a test platform with limited bandwidth and a response function that is not perfectly linear.
The neuroaccelerometer was able to learn the benchmark tasks, in spite of these non-ideal characteristics of the test system, demonstrating at the same time its robustness and its usefulness as a device that can be easily adapted to real-world systems.

\section{Design}\label{sec:design}
A necessary property of physical RC is the ability to map their input signals into a high-dimensional state, via non-linear dynamics \cite{lukosevicius_reservoir_2009}.
This mapping allows signals that are originally not linearly separable to be represented in a space where they can be processed by linear models.
In this study, the non-linear expansion of the input results from the dynamical response of a clamped-clamped beam oscillating at large amplitudes \cite{ekinci_nanoelectromechanical_nodate,harrie_a.c._tilmans_micro_1992}.
We have shown previously \cite{dion_reservoir_2018} that this dynamical response could be exploited to achieve significant neuromorphic computational capabilities, in a very small and energy efficient device.
In this work, we leverage the mechanical nature of the clamped-clamped beam computing system by coupling it to a suspended proof mass that implements the sensing functions of the neuromorphic MEMS.

The neuroaccelerometer thus comprises two principal mechanical elements: the non-linear oscillating beam, which has a high natural frequency (section \ref{sec:beam}); and a larger suspended inertial mass with a much lower natural frequency, designed to be sensitive to external accelerations (section \ref{sec:proof_mass_design}).
When in operation, a pump voltage applied to the inertial mass induces an electrostatic force over the beam, driving it near resonance with large displacements, in its non-linear regime.
External accelerations displace the inertial mass, thus modulating the amplitude of the driving force over the beam and consequently the beam oscillation amplitude.
The displacement of the beam is measured with piezoresistive strain gauges.
The signal from the gauges is digitized, delayed and fed back to the pump voltage, in a scheme described in section \ref{sec:RC_methods} that is useful to increase the computational power of simple dynamical systems, at the cost of reduced processing speed.

\subsection{Suspended Inertial Mass}\label{sec:proof_mass_design}
The suspended inertial mass consists of a relatively large central piece connected to a fixed substrate by compliant springs.
The motion of this inertial mass under inertial forces (external accelerations) can be well approximated by  a mass-spring-damper system.
The static sensitivity is given by \cite{yazdi_micromachined_1998}
\begin{equation}
\frac{x}{a} = \frac{m}{k} = \frac{1}{\omega^{2}_{0}},
\label{eq1}
\end{equation}
where $x$ is the mass displacement, $a$ is the external acceleration, $m$ is the mass, $k$ is the elastic constant of the suspension, and $\omega_{0}$ is the system resonance frequency. Note that the sensitivity can be increased by reducing the resonance frequency; this, however, also reduces the bandwidth of the sensor.
\begin{figure}[b]
\centering
\includegraphics[width=3.3in]{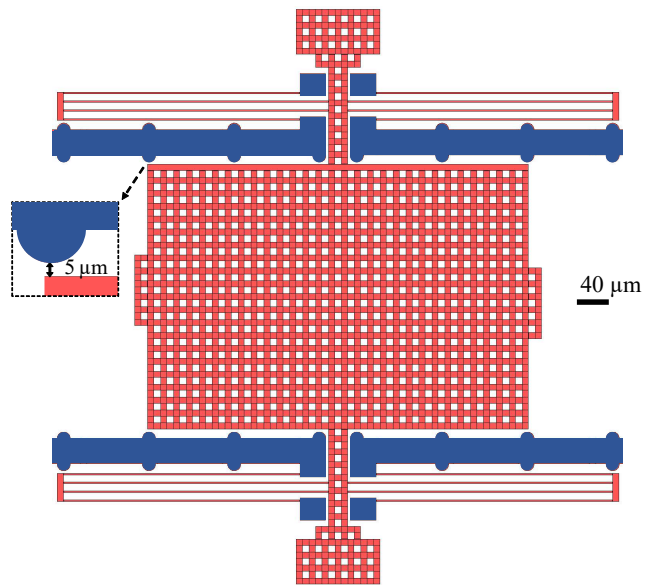}
\caption{Inertial mass suspended by a pair of accordion springs. The mass only moves in the vertical direction and its maximum displacement is 5 $\mu$m, the gap it forms with the bumpers. Blue color indicates the features that are fixed to the substrate, such as anchors and electrical traces, while structures in red are free to move.}
\label{Acc}
\end{figure}
Figure \ref{Acc} shows the suspended inertial mass composed of a 590 x 410 $\mathrm{\mu m^{2}}$ central rectangle attached to two symmetric T shaped structures.
Both the rectangular and the T shaped parts are fully perforated by 10 x 10 $\mathrm{\mu m^{2}}$ holes to facilitate fabrication. The suspension consists of a pair of 2-stage folded accordion springs that allows motion in a direction $y$. Each accordion spring possesses 4 longer members connected to the inertial mass and 4 shorter members connected to the anchors, which are fixed to the substrate. The elastic constant of the accordion spring, $k_{Acc}$, can be estimated using
\begin{equation}
k_{Acc} = \frac{4Ewt^{3}}{L_{l}^{3} + L_{s}^{3}},
\label{stiffness}
\end{equation}
where $E$ is the silicon Young's modulus, $t$ is the thickness of the members, $w$ is the width of the members, and $L_{l}$ and $L_{s}$ are the lengths of the longer and the shorter members respectively. Considering $E = 125$ GPa, $t = 2$ $\mu$m, $w = 50$ $\mu$m, $L_{l} = 410$ $\mu$m, and $L_{s} = 366$ $\mu$m, eq. \ref{stiffness} results in $k_{Acc}=1.7$ N/m, and since the two accordion springs are associated in parallel, the suspension elastic constant is $2k_{Acc}=3.4$ N/m. The 5 $\mu$m gap between the bumper and the inertial mass limits the suspension force to a maximum value of 17 $\mu$N, which corresponds to a maximum acceleration of 60g, assuming g = 9.8 $\mathrm{m/s^{2}}$ and a silicon density of 2328 $\mathrm{kg/m^{3}}$.

\begin{figure}[t]
\centering
\includegraphics[width=3.5in]{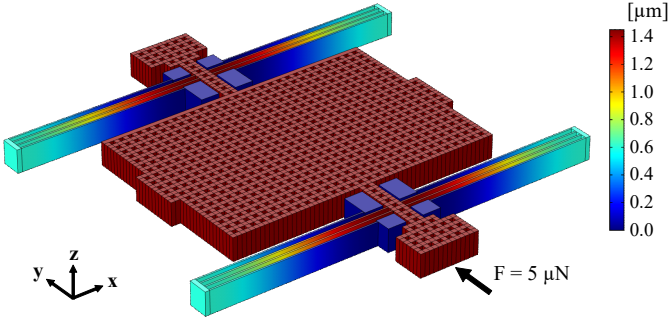}
\caption{Surface plot of the total displacement for a force of 5 $\mu$N applied in the $y$ direction. The inertial mass displaces approximately 1.5 $\mu$m in the same direction.}
\label{Acc_COMSOL}
\end{figure}

Finite element analysis (FEA) was \deleted{used}\added{performed through the solid mechanics interface of COMSOL Multiphysics\textsuperscript \textregistered \cite{noauthor_comsol_nodate}} to further simulate the static behavior of the mechanical system.
Figure \ref{Acc_COMSOL} shows the surface plot of the mass total displacement when a force of 5 $\mu$N is applied in the $y$ direction.
The ratio of the applied force to the simulated displacement gives 3.4 N/m, in agreement with the calculated spring constant.
The application of the same force in the $z$ direction results in an out-of-plane displacement of 3 nm, corresponding to an out-of-plane stiffness of 1820 N/m, approximately 535 times the in-plane stiffness.
This low cross-sensitivity is caused by the relatively large device thickness of 50 $\mu$m used in this design.
The total displacement of the mass was simulated while varying the force amplitude from 0.05 to 17 $\mu$N.
The device showed a linear behavior, with a static sensitivity of 82 nm/g.
All simulations showed negligible displacement in the $x$ direction ($<$ 1 nm), indicating sufficiently low rotation compliance.

\begin{figure}[h]
\centering
\includegraphics[width=3.5in]{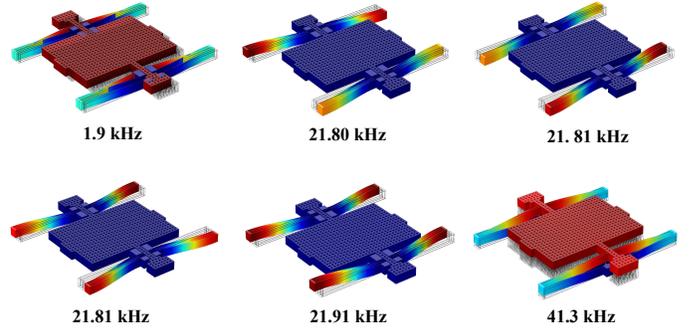}
\caption{Simulated vibration modes of the suspended inertial mass. The first 5 modes are in-plane and the last one is out-of-plane.}
\label{Vibration_modes}
\end{figure}

In addition to the static analysis, the device dynamic response was simulated (Fig. \ref{Vibration_modes}).
The eigenfrequency analysis showed in-plane natural frequencies principally around 1.9 kHz and 22 kHz.
The former corresponds to large displacements of the inertial mass in the $y$ direction.
The latter corresponds to different flexural vibration modes of the suspension springs, with negligible displacement of the inertial mass.
Furthermore, the simulation showed a first out-of-plane vibration mode at 41 kHz.
Therefore, parasitic motions are not expected to significantly affect low frequency acceleration measurements.

\subsection{Oscillating Beam}\label{sec:beam}
\begin{figure}[t]
\centering
\includegraphics[width=3.0in]{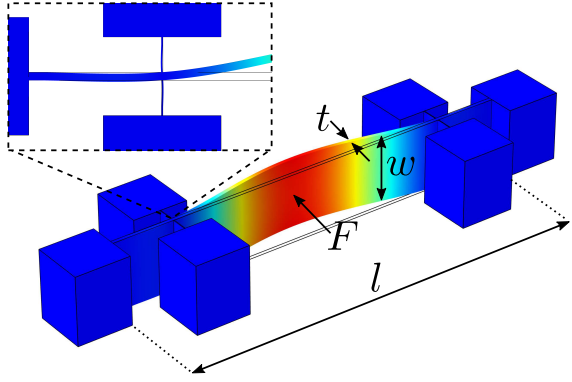}
\caption{\added{FEA model of the beam showing its first mode shape. The longitudinal strain on the piezoresistive gages allows a differential transduction of beam oscillations.}}
\label{Beam}
\end{figure}

The displacement of the oscillating beam can be approximated by the Duffing equation:
\begin{equation}
\ddot{y}+\frac{\omega_0}{Q}\dot{y}+\omega_0^2y+\beta y^3 = F(t),
\label{duffing}  
\end{equation}
where $y$, $t$, and $F$ are displacement, time, and external force per unit mass, respectively.
Dots indicate derivative with respect to time.
The beam properties, $\omega_0$, $Q$, and $\beta$, correspond to the angular natural frequency, the quality factor, and the cubic stiffness parameter, respectively.
Note that the $\beta$ term adds to the restoring spring force and introduces the non-linearity to the equation.
If $\beta$ = 0, eq. \ref{duffing} is reduced to the forced damped linear oscillator. 
In the case of a clamped-clamped beam (Fig. \ref{Beam}), the value of $\beta$ can be approximated by \cite{postma_dynamic_2005}
\begin{equation}\
   \beta = \frac{E}{18\rho} \left(\frac{2\pi}{l}\right)^4,
\label{beta_formula}
\end{equation}
where $l$ is the beam length and $\rho$ is its density.
For the \deleted{clamped-clamped} beam \added{shown in Fig. \ref{Beam}, a reduced effective beam length can be used in eq. \ref{beta_formula} in order to model the influence of piezoresistives gauges on the mode shape of the beam.}\deleted{,} 
\added{In this geometry,} $\beta$ is positive, which leads to an increase of the beam stiffness with displacement (hardening).
Eq. \ref{beta_formula} further indicates the geometric nature of this non-linearity as it depends on the beam length.
When oscillating at large amplitudes, short beams stretch significantly more than long beams, which leads to the introduction of a larger non-linear restoring force term in eq. \ref{duffing}. 
A characteristic of Duffing oscillators is the abrupt change of the oscillation amplitude for small shifts of force amplitude or driving frequency near the oscillator natural frequency.

In addition to the non-linearity, the natural frequency and quality factor also influence the beam computing performance.
The combination of \added{a} high natural \deleted{frequencies}\added{frequency} (larger than 10$^{5}$ Hz) and relatively low quality \deleted{factors}\added{factor} ($\sim$100) leads to \added{a} higher processing \deleted{speeds}\added{speed} \added{$\tau^{-1} = \frac{\pi f_0}{NQ}$ \cite{dion_reservoir_2018}, where $N$ is the number of virtual nodes (see section \ref{sec:RC_methods}).}
The optimization of the beam properties led to a beam length $l$ of 300 $\mu$m and a \deleted{width $w$}\added{thickness $t$} of 3 $\mu$m. The beam \deleted{thickness $t$}\added{width $w$} of 50 $\mu$m was defined by the fabrication technology.
Using eq. \ref{beta_formula} \added{with an effective beam length of 280 $\mu$m}\deleted{, the determined beam length} yields an expected $\beta$ value of 1.0$\times$10$^{24}$ (Hz/m)$^2$, which \added{fits our experimental data well and} provides rich enough computing dynamics for the desired application.
Simulations of the defined beam show a natural frequency $f_0 = \omega_0/(2\pi)$ of 484 kHz.
The squeeze film effect is expected to dominate viscous damping since the gap between the proof mass and the beam is small compared to the beam width.
The beam quality factor is estimated using \cite{bao_analysis_2005}
\begin{equation}
   Q = \frac{\rho t d^3 \omega_0  }{\mu w^2},
\label{Q}
\end{equation}
where $\mu$ is the dynamic viscosity of air and $d$ is the gap distance of 8 $\mu$m.
For $\mu$ = 1.8x$10^{-5}$ Pa$\cdot$s, the quality factor of the beam is 241.

\subsection{Inertial Mass Coupled to the Oscillating Beam}
\begin{figure}[t]
\centering
\includegraphics[width=3.5in]{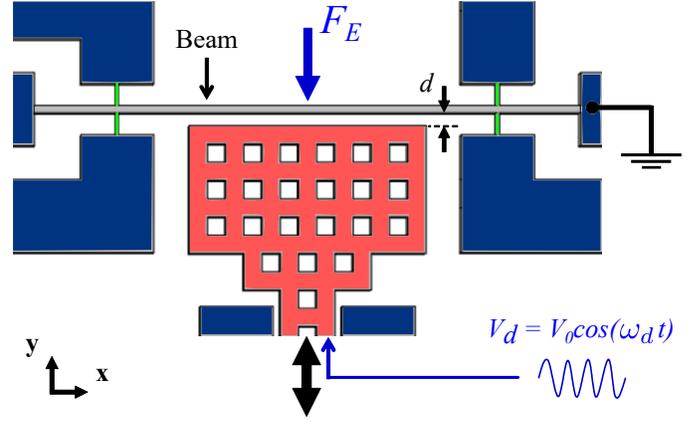}
\caption{Schematics of the coupling between the inertial mass (red) and  the beam (gray). The motion of the mass modulates the force on the beam, which oscillates near its natural frequency and in its non-linear regime. Piezoresistive gauges (green) measure the variation of the beam oscillations amplitude.}
\label{beam_plus_mass}
\end{figure}

The micro-fabricated device enables the electrostatic coupling of the suspended mass to the oscillating beam.
The beam is located next to the inertial mass, forming a gap $d$ of 8 $\mu$m with its T shaped structure, as can be seen in Fig. \ref{beam_plus_mass}.
An electrical signal is applied on the inertial mass through its anchors in order to generate an attractive electrostatic force on the beam.
This driving signal can be written as $V_d=V_{0}\cos(\omega_d t)$, where $V_{0}$ is the voltage amplitude and $\omega_{d}$ is the angular driving frequency.
The electrostatic force $F_{E}$ is estimated using the infinite parallel plate approximation, with
\begin{equation}
F_{E} \simeq \frac{\varepsilon_0 A}{2d^2} V_0^2 \cos^2(\omega_d t) = \frac{\varepsilon_0 A V_0^2}{4d^2} \left[ 1 + \cos(2 \omega_d t) \right],
\label{FE}
\end{equation}
where $\varepsilon_0$ is the dielectric constant of air and $A$ is the capacitor effective area \added{of 130 $\mu$m $\times$ 50 $\mu$m}.
From eq. \ref{FE}, one concludes that the force is proportional to $V_{0}^2/d^{2}$ and that the beam is driven at twice the driving frequency.

In experiments, $\omega_{d}$ is adjusted so that $2\omega_{d}$ is approximately the natural frequency of the beam, leading to large oscillation amplitudes, which are measured by piezoresistive gauges attached to the beam  extremities, as shown in Fig. \ref{beam_plus_mass}.
$V_{0}$ is then fine tuned to reach the beam non-linear regime, which is detected by the shift in the resonance frequency due to the stiffening of the beam.
External accelerations stimulate the motion of the inertial mass, which opens or closes the gap $d$ with the beam resulting in a change of the electrostatic force.
As a result, the external accelerations modulate the beam oscillations amplitude, albeit non-linearly.
Therefore, the described device is sensitive to external accelerations and produces non-linear outputs.

Note that since the natural frequency of the suspended mass is more than two orders of magnitude smaller than that of the beam, the inertial mass motion is not dynamically amplified when the beam is driven at its resonance.
Nevertheless, the mass equilibrium position is shifted due to the constant force term in eq. \ref{FE}, $\varepsilon AV_0^2 / (4d^2)$.
The static deflection of the beam is smaller due to its higher stiffness.

\section{Fabrication}\label{sec:fabrication}
The device was built on a (100) silicon on insulator (SOI) wafer.
The SOI was p-doped so as to increase silicon gauge factor and thus the sensitivity of the piezoresistive gauges.
The device layer, of a thickness of 50 $\mu$m, and the handle layer, both with a resistivity of (0.015 $\pm$ 0.005) $\Omega$cm, were separated by a buried oxide layer (BOX) of 4 $\mu$m.
The manufacturing flow consisted of a sequence of the following processes: photolithography, silicon etching, liberation, and metallization.

\begin{figure}[t]
\centering
\includegraphics[width=3.5in]{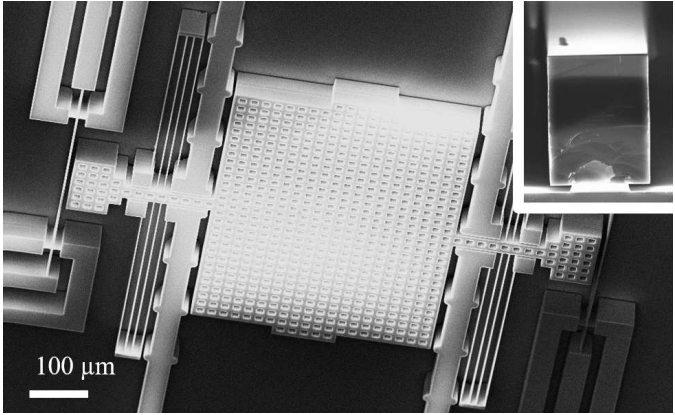}
\caption{SEM image of the MEMS neuromorphic accelerometer built on a SOI wafer. The inset shows the cross section of an anchored feature. At the end of fabrication, the wider features (anchors and electrical traces) remain connected to the substrate by the sacrificial oxide that was not etched.}
\label{SEM_final}
\end{figure}

Firstly, after dicing and proper cleaning, the substrate was spin-coated with positive photoresist AZ 9245 at 3800 RPM for 1 minute to obtain a 4.5 $\mu$m thick film on the top of the device layer.
After a soft bake of 30 minutes at 110$^\circ$C, the photoresist was exposed through the photomask to a UV dose of 200 mJ/cm$^2$ and then developed for approximately 5 minutes.
Next, the patterned dice underwent a deep reactive-ion etching step (DRIE), in which the device layer was etched 50 $\mu$m down, at roughly 90$^\circ$, reaching the BOX layer.
Then, in a procedure similar to \cite{fukuta_vapor_2003}, the dice was fixed to a silicon support wafer of 6 inches, which was flipped and placed over a Teflon dish containing 50 ml of HF 49$\%$.
A 500 W lamp was used to heat the back of the wafer to evaporate the HF to etch the oxide.
The temperature at the wafer was tuned to 40$^\circ$C since lower temperatures may condense the HF and higher temperatures would reduce the etch rate.
After 3 etching periods of 4 minutes each, the inertial mass, beam, and piezoresistive gauges were free to move, while wider features (anchors and the electrical traces) were still connected to the handle through the remaining oxide.
Figure \ref{SEM_final} shows the MEMS device after the liberation, the inset showing the cross section of a feature fixed to the handle via the oxide.
Etching cycles of at most 4 minutes were necessary since longer etching periods resulted in stiction caused by the condensation of the HF vapor.

Finally, a laser cut stainless steel hard mask was placed over the dice so that only the electrical traces and the bonding pads were exposed, i.e. not covered by the mask.
Then, a 5 nm thick film of chrome and a 200 nm thick film of gold were evaporated on the device.
The BOX kept the device and the handle layer (partially covered with gold) electrically isolated.

\section{Characterization}\label{sec:charac}

\subsection{Doubly Clamped Beam}
%actuation
The characterization of the beam dynamical response is crucial since the beam is the source of non-linearity used by the RC.
In order to test the beam, a sinusoidal driving signal of amplitude $V_0$ was applied on the inertial mass, which was kept still (no acceleration), functioning as a fixed drive electrode.
The same driving signal was applied on the handle layer in order to prevent electrostatic out-of-plane forces induced by a charged substrate (these forces could lead to the pull-in of the inertial mass onto the substrate).

%readout
Oscillations of the beam were measured via the piezoresistive strain gauges in a differential configuration.
This allowed cancellation of the large electrical feed-through of the driving signal, which was mainly caused by parasitic capacitance present on the device.
This capacitance resulted from the thin oxide layer that separated the polarized handle layer and the large wire-bonding pads used for the readout.
The $\sim$1 mV displacement signal stemming from the elongation of the strain gauges, which was one order of magnitude smaller than the feed-through signal, was amplified with a total gain of 760 and bandpass-filtered around $2\omega_d$ (passband width of 80 kHz) to mitigate noise and attenuate any leftover feed-through at $\omega_d$.
The oscillating signal was then demodulated by an envelope detector and sampled by a 16 bit analog-to-digital converter (ADC).

\begin{figure}[t]
  \centering
  \includegraphics[clip=true,trim=0 0 0 0, width=1\linewidth]{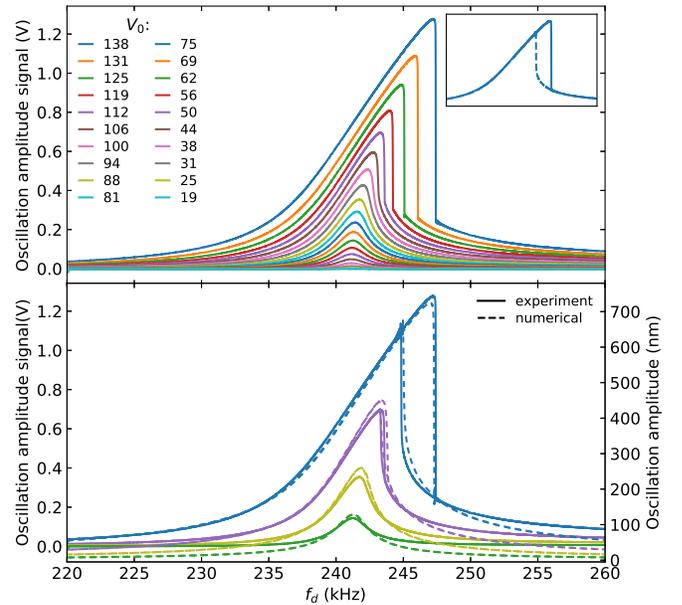}
  \caption{\added{Measured oscillation amplitude signal (top) and comparison with the oscillation amplitude of the numerical model (bottom) as a function of the driving frequency $f_{d}=\omega_d/2\pi$ for different values of $V_0$. As $V_0$ increases, the shift of the resonance frequency and appearance of jumps indicate the onset of non-linearity. The inset shows the hysteresis at $V_0=138$ V for a sweep increasing (solid line) and decreasing (dashed line) in frequency.}}
  \label{fig:beamFreq}
\end{figure}

%freq sweep
Sweeping the driving frequency $\omega_d$ from low to high values and varying $V_0$ produced the curves shown in Fig. \ref{fig:beamFreq}.
For low values of $V_0$, the beam behaved as a harmonic oscillator with a natural frequency of 482.2 kHz (twice $\omega_d/2\pi = 241.1$ kHz) and a quality factor $Q$ of 145 $\pm$ 10.
The measured natural frequency is in good agreement with the simulated value of 484 kHz; however, the measured quality factor is significantly smaller than the calculated value of 241.
This indicates that in addition to squeeze film damping, other loss mechanisms were active in the system.
As $V_0$ was increased, the peaks of the curves shifted to higher frequencies and, above the critical value of $V_0 = 106$ V, an abrupt change (jump) to lower displacement amplitudes occurred at the peak frequency.
This behavior was associated with a hysteretic response.
If the driving signal was instead swept from high to low frequencies, the curve jumped to a higher amplitude at a frequency that was lower than the sweep up peak frequency (Fig. \ref{fig:beamFreq}, inset).
This is typical of a non-linear hardening Duffing oscillator ($\beta > 0$).
In Duffing oscillators, the steady-state solution, for forcing frequencies between the two jumps (up and down), is multivalued with two stable branches and an unstable one.

\begin{figure}[t]
  \centering
  \includegraphics[clip=true,trim=0 0 0 0, width=1\linewidth]{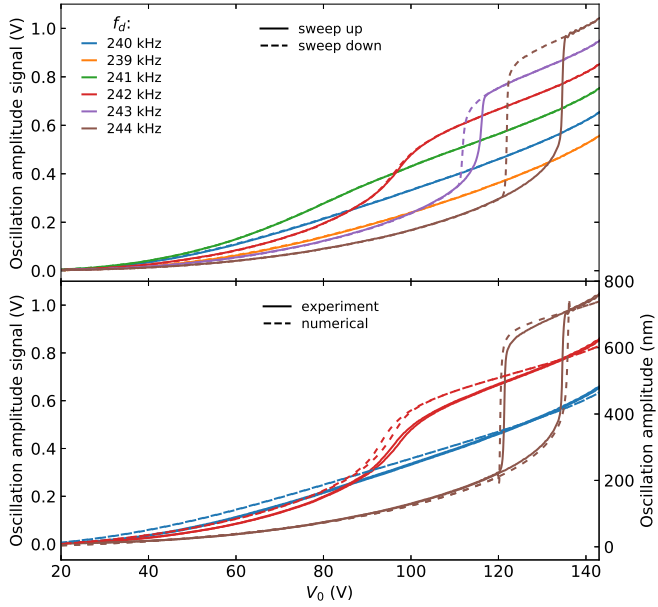}
  \caption{\added{Measured oscillation amplitude signal (top) and comparison with the numerical model (bottom) as a function of the driving voltage amplitude $V_0$ for different values of $f_d$.}}
  \label{fig:beamAmp}
\end{figure}

%amp sweep
Figure \ref{fig:beamAmp} shows the beam oscillation amplitude signal as a function of $V_0$ for different values of $\omega_d$.
Similarly to Fig. \ref{fig:beamFreq}, the beam displacement became more non-linear as the driving frequency was increased (near $\omega_d = \omega_0/2$) and for $\omega_d/2\pi > 242$ kHz an hysteretic jump shifted the curves to higher (lower) amplitudes for a sweep up (down).
Note that the higher the driving voltage frequency $\omega_d$, the larger was the jump or change of the oscillation amplitude; and therefore, the higher was the non-linearity. 
The square law behavior apparent in these curves, not described by the Duffing equation, is due to the quadratic dependence of the force with respect to voltage (eq. \ref{FE}) and also contributed to the non-linearity of the system.

\added{Numerically solving equation \ref{duffing} using equations \ref{beta_formula}, \ref{Q} and \ref{FE} while accounting for the electrostatically induced change in equilibrium position of the inertial mass for every new value of $V_0$ (i.e. $d=d(V_0)$) produced the dashed curves in the bottom panel of figures \ref{fig:beamFreq} and \ref{fig:beamAmp}.}

\subsection{Inertial Mass Response to Accelerations}
The motion of the inertial mass was sensed by measuring the change of the piezoresistive signal caused by the variation of the gap distance $d$ (see eq. \ref{FE}).
As discussed previously, the application of external acceleration in the $y$ direction leads to the displacement of the inertial mass, which closes or open the gap with the beam.

Acceleration signals were applied on the device using a PID-controlled electrodynamic shaker.
The device was mounted near a calibrated integrated electronics piezo-electric (IEPE) accelerometer used for the control loop.
For this characterization, the beam was operated in its linear regime, at $\omega_d/2\pi$ = 241.1 kHz and $V_0$ = 75 V, to obtain an accurate assessment of the inertial mass response.
The piezoresistive signal was processed as before to produce an electrical signal proportional to the beam oscillation amplitude.

\begin{figure}[t]
  \centering
  \includegraphics[clip=true,trim=0 0 0 0, width=1\linewidth]{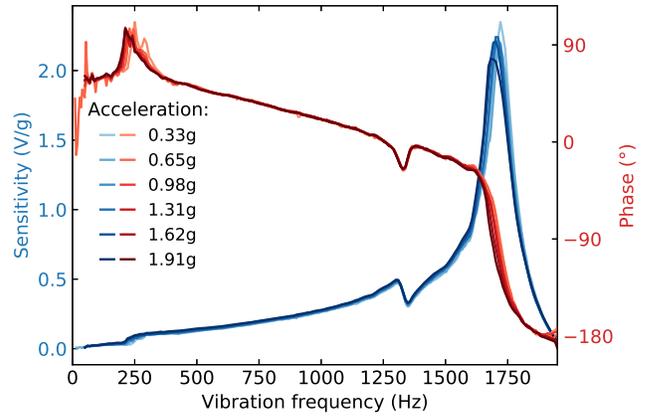}
  \caption{Inertial mass sensitivity as a function of the vibration frequency for different acceleration amplitudes. The peaks near 1.7 kHz indicate the resonance frequency of the inertial mass.}
  \label{fig:pmFreq}
\end{figure}

Figure \ref{fig:pmFreq} shows the device sensitivity for different sinusoidal acceleration amplitudes, sweeping the shaker vibration frequency from 10 Hz to 2 kHz.
The graph shows a plateau from 250 Hz to 1.3 kHz, where the sensitivity was on the order of 100 mV/g, independent of the acceleration amplitude.
The first vibration mode of the inertial mass occurred at \mbox{$(1706 \pm 5)$} Hz, with a quality factor of $19 \pm 2$.
The measured natural frequency of the suspended mass showed good agreement with the simulated value of 1865 Hz, less than 10\% error.
This difference is likely due to the observed 10-15\% reduction of the spring thickness caused by the occurrence of notches in the end of the DRIE.
\added{While the features appearing around 250 Hz and 1300 Hz are not fully understood, possible causes include vibration modes in the packaging of the MEMS device, rotational modes of the shaker and vibration or displacement of other structures present on the device (multiple neuroaccelerometers are patterned on a single chip for convenient testing). 
Despite this non-ideal sensor response, the device was able to perform non-trivial computations on sensed inputs in the neuroaccelerometer configuration (section \ref{sec:processing}), which underscores the robustness of such neuromorphic sensor systems.}

\section{Processing Acceleration Signals with the Neuroaccelerometer}\label{sec:processing}

\subsection{Methods}\label{sec:RC_methods}
\begin{figure}[t]
\centering
\includegraphics[width=1\linewidth]{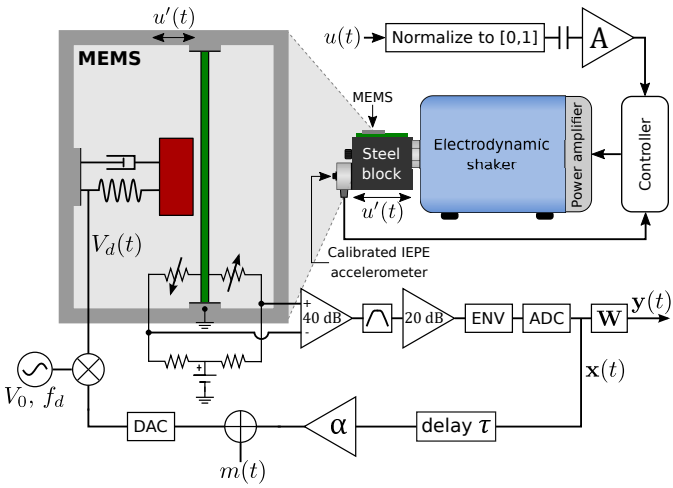}
\caption{\added{Experimental setup of the neuroaccelerometer.}}
\label{fig:setup}
\end{figure}

Neural-like processing capabilities were conferred to the neuroaccelerometer by using the non-linear oscillating beam as the single physical node in a delay-coupled RC \cite{appeltant_information_2011}. The system, schematized in Fig. \ref{fig:setup},
was essentially the same as that of our previous MEMS RC \cite{dion_reservoir_2018}, with the difference that voltage signal inputs on the fixed electrode of the MEMS RC were replaced by acceleration signals sensed by the suspended inertial mass, which in turn modulated the electrostatic forcing of the beam in a gap-closing configuration.

%masking, beam timescale
In this scenario, the sinusoidal electrostatic pump was first amplitude-modulated by a temporal mask $m(t)$ of period $\tau$, with mask values updated at a rate $\theta^{-1}$. Each sample of the mask defined a virtual node of the reservoir, and the mask values corresponded to input weights. By choosing $\theta \lesssim T$, where $T=2Q/\omega_0=(96 \pm 7) \ \mathrm{\mu}$s is the beam ring-down characteristic time, the nodes were kept from settling to a steady-state value before the mask was updated. This led to an interconnection scheme where each virtual node was coupled at least to one of its nearest neighbor: the state of a given node depended on the state of the previous node, a time $\theta$ in the past. While this mask signal could take different forms, such as different random distributions \cite{paquot_optoelectronic_2012,kuriki_impact_2018}, sinusoidal signals \cite{duport_analog_nodate}, multiple level digital signals \cite{soriano_optoelectronic_2013} and even chaotic signals \cite{nakayama_laser_2016}, we chose the most commonly used form of a random binary sequence for the sake of simplicity. In this case, the mask was composed of only two values and switched randomly between them after each interval $\theta$. Values of 0.45 and 0.7 produced adequate results for both our neuroaccelerometer and fixed drive MEMS RC \cite{dion_reservoir_2018}.

%acceleration, proof-mass timescale
For both benchmarks presented below, the input samples ($u(t)$ in Fig. \ref{fig:setup}) were scaled to an amplitude of 2g (gain $A=4$g), then held for a duration $\tau$ and output via a DAC to the shaker PID controller.
Since $\tau$ was also the RC output update period and the number of virtual nodes was $N=\tau/\theta$ with $\theta$ determined by $T$, the dynamical properties of the beam ($T$) and of the acceleration input ($\tau$) dictated the maximum available number of virtual nodes. As harder tasks typically require a larger reservoir size $N$, they also demand a faster reacting beam, i.e. lower $Q$ and higher $\omega_0$.

Memory of recent inputs, which is necessary for processing time series such as the acceleration signals used in sections \ref{sec:NARMA} and \ref{sec:parity}, was provided to the RC by a delayed feedback loop. The latter added the previous virtual node states to the current mask pattern such that each virtual node was driven by a superposition of the masked pump signal, the inertial mass physical displacement and its previous response, $\tau$ seconds earlier. The voltage applied on the inertial mass thus took the form
\begin{equation}
V_d(t) = V_0 \left[m(t) + \alpha x(t-\tau) +1 \right] \cos\left( \omega_d t \right),
\end{equation}
where $x(t)$ is the beam displacement envelope signal sampled by the ADC and $\alpha$ is the feedback gain. The output of the envelope detector was sampled at the end of each interval $\theta$ so that $N$ samples were collected per period $\tau$, yielding the vector $\mathbf{x}(k)$ containing the reservoir state at timestep $k = (t-t_0)/\tau$. Finally, the RC output vector $\mathbf{y}(k)$ was constructed from linear combinations of the virtual node states:
\begin{equation}\label{eq:RCOutput}
\mathbf{y}(k) = \mathbf{W} \mathbf{x}^\intercal(k) .
\end{equation}
$\mathbf{W}$ is a readout weight matrix with each row corresponding to the weights for a different dimension of $\mathbf{y}(k)$, and $\mathbf{y}(k)$ is a column vector. The matrix $\mathbf{W}$ was computed offline, following a training phase where, after discarding the initial transient, $M$ reservoir states were accumulated in an $M \times (N+1)$ matrix $\mathbf{X}$ such that each row contained the reservoir state $\mathbf{x}(k)$ for $k=0,1,..,M-1$, augmented with a constant bias term (needed to reproduce signals with non-zero mean). Training the readout using a ridge regression,
\begin{equation}\label{eq:training}
\mathbf{W} = \mathbf{Y}' \mathbf{X}^\intercal \left( \mathbf{X} \mathbf{X}^\intercal + \mathbf{\Gamma } \right)^{-1} ,
\end{equation}
where $\mathbf{Y}'$ is the matrix of desired outputs constructed in a similar way to $\mathbf{X}$, allowed to prevent from over-fitting by introducing a regularization matrix $\mathbf{\Gamma}$. 
Choosing $\mathbf{\Gamma} = \gamma \mathbf{I}$, where $\mathbf{I}$ is the $(N+1) \times (N+1)$ identity matrix, produced satisfactory results. The regularization parameter $\gamma$ was optimized every time a new readout weight matrix was computed by choosing the value which maximized the performance of the RC in the testing phase for the given task.

\added{The electronic circuitry shown at the bottom of Fig. \ref{fig:setup} was implemented by a custom analog front end combined with various commercially available instruments. The main obstacles to full integration of the control electronics are the high voltage drive signal and the delay loop. The former can be dealt with by miniaturizing the device: according to our model, reducing all geometrical dimensions of the oscillating beam by one order of magnitude (except the gap $d$, which only needs to be scaled by a factor of 1/5) would allow drive voltages below \mbox{10 V}. The latter, which currently necessitates analog-to-digital conversion in order to appropriately delay the feedback signal, could eventually be bypassed by coupling multiple oscillating beams in a scheme first described in \cite{coulombe_computing_2017}, which does not require external delayed feedback.}

\subsection{NARMA Benchmark}\label{sec:NARMA}
\begin{figure}[!h]
  \centering
  \includegraphics[clip=true,trim=0 0 0 0, width=1\linewidth]{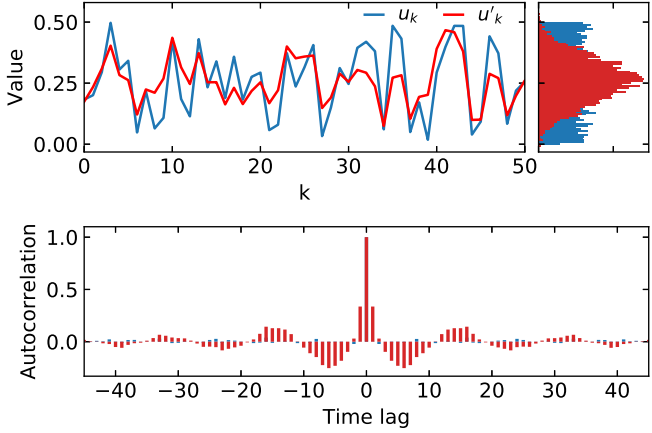}
  \caption{The uniform samples $u(k)$ used for the PID setpoint (top left) are altered by the vibration system, resulting in a different distribution (top right) of samples $u'(k)$ which are used to compute the target (eq. \ref{eq:NARMA_target}). The bottom panel shows the correlogram for the two signals.}
  \label{fig:NARMA_input}
\end{figure}

The emulation of non-linear autoregressive moving average (NARMA) models is a widespread machine learning benchmark task \cite{jaeger_adaptive_nodate,steil_backpropagation-decorrelation:_2004,atiya_new_2000,verstraeten_experimental_2007,paquot_optoelectronic_2012}. \added{This benchmark is relevant as such non-linear filters are often used for signal processing and control applications.}  An important part of its appeal stems from its (non-linear) memory requirements: its current output is a non-linear combination of many past inputs and outputs. By introducing a time-lag parameter $n$, a generalized version of its input-output relationship can be written as
\begin{equation}\label{eq:NARMA_target}
\begin{aligned}
y_n(k+1) = \ & 0.3y_n(k) + 0.05y_n(k)\sum_{i=0}^{n-1} y_n(k-i)\\
 & + 1.5u(k)u(k-n+1) +0.1 ,
\end{aligned}
\end{equation}
where $u(k)$ is typically an i.i.d. uniform random variable \cite{jaeger_adaptive_nodate,atiya_new_2000,kubota_dynamical_2019} over [0, 0.5] constituting the input sequence. 
%A tanh saturation is often applied to the RHS of equation \ref{eq:NARMA_target} in order to improve its stability, especially for higher values of n [ref rodan 2011, goudarzi2014].
For this study, this random sequence was input, after the scaling procedure described in section \ref{sec:RC_methods}, on the electrodynamic shaker PID controller as a voltage signal, resulting in an acceleration signal $u'(k)$ that was a distorted version of the original input, as shown in Fig. \ref{fig:NARMA_input}. Indeed, because of the non-ideal mechanical response of the shaker, the original uniform distribution became gaussian-like and samples separated by less than 40 timesteps became correlated.
In order to exempt the RC from also inverting the shaker transfer function, which would constitute a harder task, the actual acceleration signal measured with the reference accelerometer was sampled, rescaled and offset so that its amplitude and mean were similar to those of $u(k)$, then used instead of $u(k)$ to compute the target (i.e., $u'(k)$ was used instead of $u(k)$ in eq. \ref{eq:NARMA_target}). The RC was then trained (eq. \ref{eq:training}) using the first $M = 4000$ samples of $u'(k)$. For this task, each dimension of $\mathbf{y'}$ contained the expected NARMA$_n$ output for a different value of the time-lag parameter $n=2,3,..,20$:
\begin{equation}
  \begin{aligned}
    \mathbf{y'}(k) &= \begin{bmatrix}
           \mathrm{NARMA}_{2}(k) \\
           \mathrm{NARMA}_{3}(k) \\
           \vdots \\
           \mathrm{NARMA}_{20}(k)
         \end{bmatrix} .
  \end{aligned}
\end{equation}
Following the training phase, the RC was tested by collecting the next 400 samples of the RC output $\mathbf{y}(k)$ and comparing it to $\mathbf{y'}(k)$ via the normalized root mean squared error for each dimension:
\begin{equation}
\mathrm{NRMSE}_n = \sqrt{\frac{\mathbb{E} \left[ \left( \mathbf{y'}_n(k) - \mathbf{y}_n(k) \right)^2 \right]}{\sigma ^2 (\mathbf{y'}_n(k))}} ,
\end{equation}
where $\mathbb{E}$ denotes the expected value over $k = 4000, 4001,..,4399$ and $\sigma ^2 (\mathbf{y'}(k))$ is the variance of the target. This metric was then used as a criterion for adjusting the RC hyperparameters, yielding optimized values of $(V_0, f_d, \alpha, \theta, \gamma) = ($135 V, 245 kHz, 1.2, 50 \textrm{$\mu$}s, $10^{-3}$ V$^2$).

\begin{figure}[t]
  \centering
  \includegraphics[clip=true,trim=0 0 0 0, width=1\linewidth]{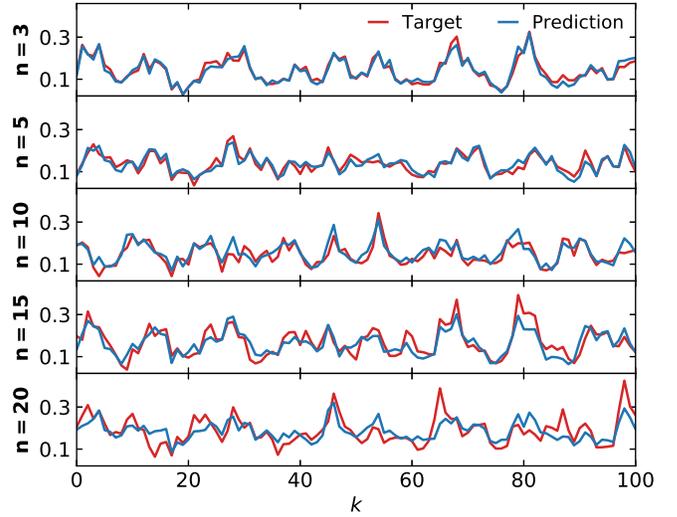}
  \caption{First 100 timesteps of the RC output (eq. \ref{eq:RCOutput}) and of the target (eq. \ref{eq:NARMA_target}) in the testing phase for the NARMA task with $n=3,5,10,15,20$.}
  \label{fig:NARMA_results}{}
\end{figure}

\begin{figure}[t]
  \centering
  \includegraphics[clip=true,trim=0 0 0 0, width=1\linewidth]{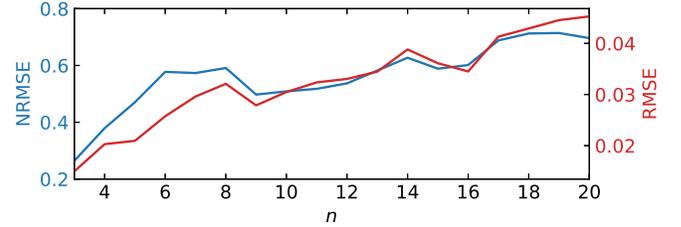}
  \caption{NRMSE and RMSE of the RC emulation of eq. \ref{eq:NARMA_target} as a function of the time-lag parameter $n$.}
  \label{fig:NARMA_NRMSE}
\end{figure}

%discuss results
Figure \ref{fig:NARMA_results} shows the RC output overlaid on the target waveform for $n=3,5,10,15,20$. The reproduction of the target became worse as the memory requirement of the task increased between $n=3$ and $n=20$, as shown in Fig. \ref{fig:NARMA_NRMSE} where the testing error is shown against the time-lag parameter. The root mean squared error (RMSE) did not exactly follow the NRMSE due to different variances for the different dimensions of $\mathbf{y'}(k)$. At its current state of optimization, the neuroaccelerometer did not reach the performance of some noiseless software RC (NRMSE values close to 0.1 \cite{ferreira_approach_2013,jaeger_adaptive_nodate} for $n=10$, compared to a value of 0.5 for the neuroaccelerometer), but performed similarly to other hardware \cite{paquot_optoelectronic_2012} and software \cite{verstraeten_experimental_2007} RC. Noise inherent to physical systems hinders their memory capacity \cite{soriano_delay-based_2015}, making processing time series with long memory requirements more challenging.

\subsection{Parity Benchmark}\label{sec:parity}
The parity benchmark allows a straightforward comparison of the neuroaccelerometer computing capabilities with other systems \cite{dion_reservoir_2018,coulombe_computing_2017,phua_parallel_2003,adhikari_memristor_2012,bertschinger_real-time_2004,nakajima_exploiting_2014,koprinkova-hristova_introduction_2015}. \added{It is a non-linear pattern classification task requiring a memory of past inputs which makes it well suited for RC evaluation, as a prototype for non-trivial signal recognition tasks.} A random binary stream $u(t)$ was fed to the shaker controller as a voltage signal, resulting in a zero-mean, 2g amplitude acceleration waveform input to the MEMS. Since the electrodynamic shaker has a maximum travel distance of 13 mm, long sequences of successive identical values in the input would have been heavily distorted. In order to mitigate this, input bits were flipped when the input stream exceeded 3 successive bits without switching. 
\begin{figure}[t]
  \centering
  \includegraphics[clip=true,trim=0 0 0 0, width=1\linewidth]{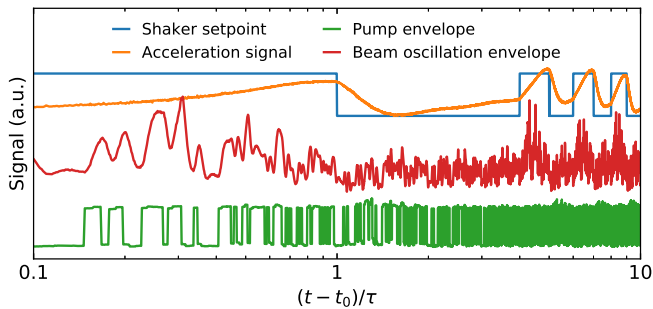}
  \caption{Typical behavior of the different signals involved in computing the parity function with the RC, showing different timescales on a logarithmic scale.}
  \label{fig:parity_waveforms}
\end{figure}
Figure \ref{fig:parity_waveforms} shows that the acceleration signal was still distorted with respect to the controller setpoint due to the non-ideal shaker response, especially for signals with a large bandwidth such as the step-wise input for the parity task. Showing the signals on a logarithmic scale allows the observation of the different timescales involved in the system. On the shorter timescales of order $T\sim\theta=50 \ \mathrm{\mu}$s, the electrostatic pump envelope alternated between the mask values, and the beam oscillation amplitude was a non-linear function of this signal. The right half of Fig. \ref{fig:parity_waveforms} shows that the beam response was also modulated by the slowly evolving acceleration signal (through the inertial mass displacement, not shown), of characteristic time $\tau=N\theta=5$ ms.

Following this excitation pattern, the neuroaccelerometer had to compute the parity (without delay) of order $n=1$ to $n=6$:
\begin{equation}
  \begin{aligned}
    \mathbf{y'}(k) &= \begin{bmatrix}
           P_{1}(k) \\
           P_{2}(k) \\
           \vdots \\
           P_{6}(k)
         \end{bmatrix},
  \end{aligned}
\textrm{where} \ P_{n}(k) = \prod_{i=0}^{n-1} u\left(k-i\right),
\end{equation}
where the input $u(k)$ was a unit-amplitude version of the signal sent to the PID controller.
\begin{figure}
\centering
\includegraphics[width=3.5in]{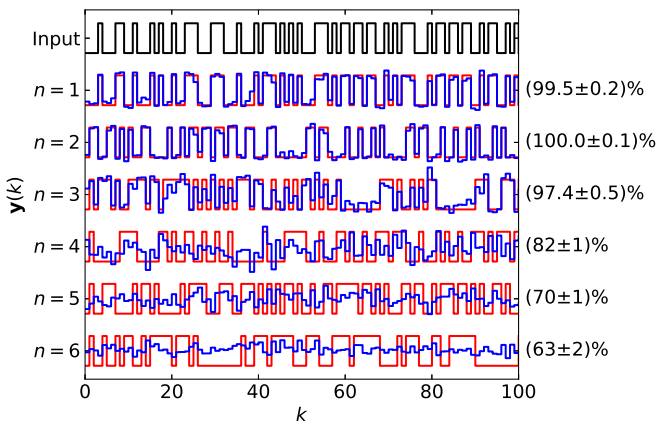}
\caption{Following excitation of the RC with the input bit stream (black), the RC output waveforms for the $P_1$ to $P_6$ tasks (blue) are thresholded and compared to the target waveforms (red) to produce the success rates shown on the right.}
\label{fig:parity}
\end{figure}
As for the NARMA benchmark, the training sequence lasted 4000 samples, but the testing phase was longer at 2000 samples in order to be more precise on the performance metric, which was computed by thresholding the RC output at 0 and comparing the sign of the resulting bit stream with that of the target. This last step produced the success rates shown beside their respective waveforms in Fig. \ref{fig:parity}, with the half-width of the 95\% confidence Agresti-Coull interval \cite{agresti_approximate_1998} used as the uncertainty. Hyperparameters for Fig. \ref{fig:parity} were $(V_0, f_d, \alpha, \theta, \gamma) = ($135 V, 245 kHz, 0.7, 50 \textrm{$\mu$}s, 0.005 V$^2$). The RC output for $n=1$, which is the identity function of the PID setpoint $u(t)$, showed that the neuroaccelerometer was able to accurately invert the non-linear response of the shaker.
As was the case for the NARMA benchmark, the time traces became more noisy when the memory requirement of the task was increased from $n=1$ to $n=6$.
While thresholding them yielded lower success rates than it did for our fixed drive MEMS RC for $n \geq 3$ \cite{dion_reservoir_2018}, the latter had direct access to the undistorted input as a voltage signal, making the task easier.

\section{Conclusion}
This paper demonstrated the design, fabrication, and validation of a MEMS accelerometer with built-in computing capabilities.
The concept of exploiting the non-linearity of a clamped-clamped micro beam to emulate a reservoir comput\deleted{ing}\added{er} had been previously simulated \cite{coulombe_computing_2017} and experimentally demonstrated \cite{dion_reservoir_2018}.
In this study, an accelerometer was coupled to the computing beam in order to integrate sensing and computing in a single MEMS.
This is the first experimental demonstration of such a highly integrated device, which represents a new class of MEMS devices.
Through integration, these devices have the potential to be much smaller, faster, and more energy efficient than conventional combined control systems comprising sensors and separate electronic controller units.
In addition, the computing functions were implemented via a neuromorphic system that has many of the beneficial features of machine learning algorithms implemented in software.
The same neuromorphic MEMS can be trained to implement many different data processing tasks, such as classification (as demonstrated with the parity benchmark) and the implementation of complex non-linear functions (as demonstrated with the NARMA benchmark).
This training characteristic could be useful to simplify the design of control systems, to increase the robustness of applications, and to facilitate the adaptation of systems to changing environments (e.g. using continuous unsupervised training).
Machine learning systems, including RC, have been shown to frequently offer powerful generalization capabilities, with errors on validation data that are not much larger than errors on training data.
Such generalization capabilities could be especially useful to increase the robustness of control systems.

Neuromorphic MEMS could be used for applications with strict constraints on volume, weight, response time or energy consumption.
These include many autonomous and robotic applications, as well as mobile and wearable devices.
Neuromorphic MEMS could also be especially useful in distributed sensor networks (e.g. the Internet of Things), to limit the amount of data transmitted to a central processing unit by providing significant computing power ``at the edge'', to transmit only data corresponding to specific patterns.
The technological relevance of neuromorphic MEMS will likely be enhanced by future work that could focus on integrating neuromorphic computing capabilities to other MEMS sensors, and on increasing the computational power of the MEMS RC.
Increasing the number of physical nodes in the reservoir network (multiple, coupled resonant non-linear structures) could increase the device processing speed,
leading to hybrid networks formed by physical nodes and virtual nodes (through multiplexing in time).
Such MEMS could have simpler drive electronics, to eventually reach the full technological benefits of neuromorphic MEMS sensors.

% if have a single appendix:
%\appendix[Proof of the Zonklar Equations]
% or
%\appendix  % for no appendix heading
% do not use \section anymore after \appendix, only \section*
% is possibly needed

% use appendices with more than one appendix
% then use \section to start each appendix
% you must declare a \section before using any
% \subsection or using \label (\appendices by itself
% starts a section numbered zero.)
%

%\appendices
%\section{Proof of the First Zonklar Equation}
%Appendix one text goes here.

% you can choose not to have a title for an appendix
% if you want by leaving the argument blank
%\section{}
%Appendix two text goes here.

% use section* for acknowledgment
\section*{Acknowledgment}

The authors thank the Ministère de l’économie, de la science et de l’innovation du Québec and the Natural Sciences and Engineering Research Council of Canada for financial support.

% Can use something like this to put references on a page
% by themselves when using endfloat and the captionsoff option.
\ifCLASSOPTIONcaptionsoff
  \newpage
\fi

% trigger a \newpage just before the given reference
% number - used to balance the columns on the last page
% adjust value as needed - may need to be readjusted if
% the document is modified later
%\IEEEtriggeratref{8}
% The "triggered" command can be changed if desired:
%\IEEEtriggercmd{\enlargethispage{-5in}}

% references section

% can use a bibliography generated by BibTeX as a .bbl file
% BibTeX documentation can be easily obtained at:
% http://mirror.ctan.org/biblio/bibtex/contrib/doc/
% The IEEEtran BibTeX style support page is at:
% http://www.michaelshell.org/tex/ieeetran/bibtex/
\bibliographystyle{IEEEtran}
% argument is your BibTeX string definitions and bibliography database(s)
\bibliography{IEEEabrv,References}

\end{document}